# Topological surface states of $Bi_2Se_3$ with the coexistence of Se vacancies

**Binghai Yan**[1], **Delin Zhang**[2], **Claudia Felser**[1,2]

[1] Johannes Gutenberg-Universität Mainz, Staudingerweg 9, 55128 Mainz, Germany

[2] Max-Planck-Institute für Chemische Physik fester Stoffe, Nöthnitzer Str. 40, 01187 Dresden, Germany

Although topological surface states are known to be robust against non-magnetic surface perturbations, their band dispersions and spatial distributions are still sensitive to the surface defects. Take $Bi_2Se_3$ as an example, we demonstrated that Se vacancies modifies the surface band structures considerably. When large numbers of Se vacancies exist on the surface, topological surface states may sink down from the first to second quintuple layer and get separated from the vacancies. We simulated STM images to distinguish the surfaces with Se- and Bi-terminations.

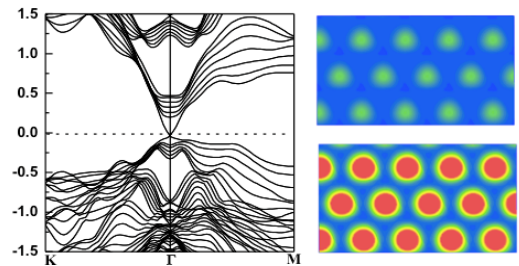

**1 Introduction** Topological insulators (TIs) accommodate metallic surface or edge states inside the bulk energy gap, which are robust against defects, impurities and disorders under the time-reversal symmetry [1,2]. Today the $Bi_2Se_3$ family [3-5] of TIs are the most extensively studied materials among many reported compounds (see ref. [6] and references therein).

Selenium vacancies are defects commonly observed in $Bi_2Se_3$. They are believed to give rise to electron doping and increase the conductivity of bulk states dramatically [7]. On the surface, Se vacancies are usually seen as clear triangular depressions in STM experiments [7,8]. Recently, numerous Se vacancies are reported to exist on the surface after the outermost Se atoms quickly desorb at room temperature. This results in an ordered Bi-terminated surface that is revealed by the LEED pattern [9]. However, most of recent theoretical studies adopted the perfect $Bi_2Se_3$ surface without defects [3,10]. Because Se vacancies usually coexist with the topological states on the surface, it is interesting to explore the interplay between them.

In this Letter, we focused on the $Bi_2Se_3$ surface with Se vacancies on the outermost atomic layer. First-principles calculations were employed to study the band structures and charge densities of the surface states. We simulated STM images of different surfaces, in order to distinguish the Bi-terminated surface from the Se-terminated one.

**2 Computational Methods** First-principles method was employed within the framework of the density-functional theory with the generalized gradient approximation [11]. We used the Vienna *ab initio* simulation package [12] with the projected augmented wave method. A thick slab model of seven quintuple layers (QLs) was adopted to simulate the $Bi_2Se_3$ surface, in which a QL stands for Se-Bi-Se-Bi-Se atomic layers in the lattice. A primitive unit cell in the *xy* plane was used to represent the Se and Bi terminated surfaces, while a 3×3 supercell was used to simulate a surface with only one Se vacancy. The top two QLs are shown in Fig.1. Vacancies were assigned to both the top and bottom surfaces of the slab, in order to keep the inversion symmetry and remove the artificial surface dipoles. The atomic positions inside the outermost QLs (except the lowest Se layer) were fully optimized, while all other atoms were fixed to the bulk positions. Based on the relaxed structures, spin-orbit coupling (SOC) was included to calculate the band structures and charge densities. We stimulated the STM images using the Tersoff-Hamann approximation [13], in which the STM cur-rent $I(r)$ is proportional to the integrated local density of states (LDOS) of the surface (at the position of the tip). At a finite STM bias $V_b$, the image is simulated in terms of integrating LDOS between $E_F$ and $E_F+V_b$,

$$I(r) \propto \int_{E_F}^{E_F+V_F} LDOS(r)dE = \int_{E_F}^{E_F+V_b} \sum_n |\psi_n(r)|^2 dE, \quad (1)$$

where $\psi_n(r)$ is the wave function with eigen energy $E_n$ ($E_F < E_n < E_F + V_b$).

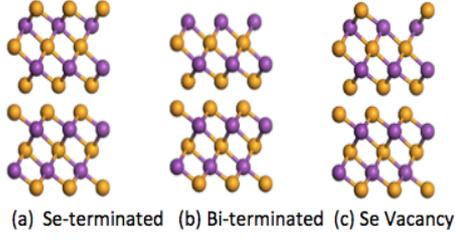

**Figure 1** The structure models of (a) a perfect surface with S-termination, (b) a Bi-terminated surface and (c) a surface hosting one Se vacancy. Only the top two QLs of the slab model are shown here. The yellow (grey) balls stand for Se atoms and purple (black) balls for Bi atoms.

**3 Results and discussion** Figure 2 shows the calculated band structures and STM images of above three surfaces. The perfect surface with Se-termination exhibits a single Dirac cone inside the bulk gap from topological surface states (TSSs), consistent with previous calculations [3] and experiments [4]. One can see a triangular lattice in the simulated STM images, in which the protrusion sites correspond to the surface Se atoms. However, several dangling bond states (topologically trivial) appear also inside the gap and hybridize with the TSSs (Figs 2b and 2c), when Se vacancies exist on the surface. When there is one Se vacancy (Fig. 2c), three Bi dangling bonds form below the missing Se site. One can find that there are three new states near the $K$ point of the surface Brillouin zone. The empty state at about 0.4 eV is mainly due to the $p$ orbitals of three Bi atoms that have dangling bonds. These two occupied states below $E_F$ are also contributed mainly by above three Bi atoms and partially by the $p$ sates from the Se atoms below these Bi atoms, i.e. the third atomic layer. The Dirac point at the $\Gamma$ point upshifts nearly to the bottom of the conduction bands, consistent with previous calculations [14]. We note that the band structure of Fig. 2c is from a supercell of periodically ordered Se vacancies. But Se vacancies may have a random distribution in experiments. Therefore, ARPES cannot observe clear dispersions of the defect bands. The Se vacancy is found to be a triangular depression in both filled state and empty state STM images (Figs 2h and 2i), agreeing well with previous experimental observation [7,8,15,16]. The existence of Se vacancy shifts the Fermi energy up close to the conduction bands, as an electron doping effect. Consequently the LDOS integral in Eq. (1) includes more (less) states for positive (negative) $V_b$. Therefore the empty (filled) state STM image looks brighter (darker) than the case of a perfect surface (Fig. 2d and 2e).

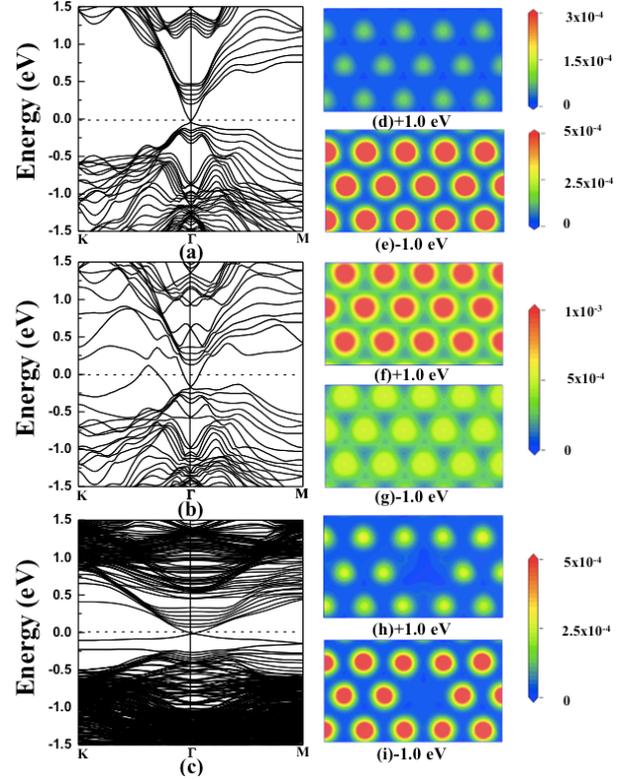

**Figure 2** Band structures for (a) perfect surface with Se-termination, (b) Bi-terminated surface, and (c) a 3×3 supercell with one Se vacancy. Simulated STM images ($I(r)$) are shown on the right of corresponding band structures. Vb= + 1.0 eV is used in the LDOS integral for empty state images and -1.0 eV for filled state images. The integrated LDOS at the position of 2 Å above the surface is adopted for STM images. Red colour represent high charge density and blue represent zero. The colour map is shown in unit of e/Å$^3$.

As the density of Se vacancies increases, the band structure becomes more complicated. When a fully Bi-terminated surface or the full vacancy coverage exists, the Dirac point returns to the top of the valence bands, similar to a Se-terminated surface (Fig. 2a). However, several trivial states emerge in the bulk energy gap. TSSs are known to disappear when turning off SOC, while these states preserve. When we project these states to the atomic orbitals, two filled states at $K$ below $E_F$ are of $p$ states of the outermost Bi and Se atoms. Two empty states at K are of $p$ states of the surface Bi layer. This is similar to the case of a single vacancy (Fig. 2c). Since the Bi-terminated surface is well ordered [9], it is possible to observe these trivial surface states as well as TSSs in ARPES experiments.

Besides the differences in band structures, the Bi-terminated surface also varies from the Se-terminated one very much in STM images. For $Bi_2Se_3$, the upper valence bands are mainly composed of Se-$p$ orbitals and the lower conduction bands are mainly of Bi-$p$ orbitals. STM detects mainly the states of the surface atoms. For a Se-terminated surface, there are more (less) Se states in the filled (empty) state image. Consequently, Se atoms are imaged brighter in Fig. 2e than those in Fig. 2d. In contrast, the empty state image of Bi-terminated surface is brighter than corresponding filled state image, since Bi orbitals mainly compose the empty bands. In this way, one can easily distinguish the surface with Bi-termination from that with Se-termination.

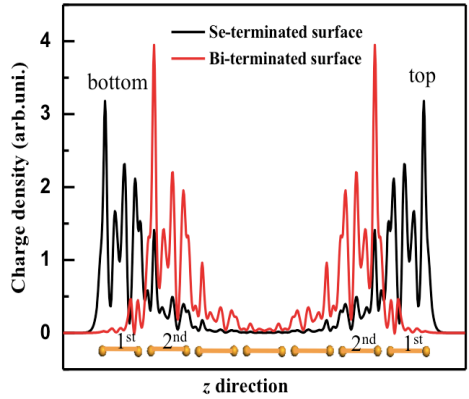

**Figure 3** Charge density distribution of the topological surface states along $z$ direction. The left side corresponds to the bottom surface of the slab model and the right side to the top surface. Seven pieces of lines indicate the QL positions of the surface slab. The balls at the ends of a line represent the end Se atoms of a QL.

As mentioned above, TSSs in a Bi-terminated surface exhibit energy dispersion quite similar to the Se-terminated one, although many dangling bonds exist and the STM images look very different. It is worth investigating the reason of the preserved Dirac cone for the Bi-terminated surface. We plotted the charge density of TSSs along the $z$ direction (perpendicular to the surface), in which the charge density is averaged in the $xy$ plane. One can see that TSSs mainly distribute inside the first QL on the perfect surface (Fig. 3), consist with previous study [17] . However, TSSs sink down into the second QL on the Bi-terminated surface. Here, TSSs are apart from the Bi dangling bonds and get away from the surface defects. Inside the second QL, TSSs show a distribution similar to that of a pristine Se-terminated surface. So it is not surprising that TSSs on Bi-terminated surface maintain the same Dirac cone type of energy dispersion (Fig. 2b) as a Se-terminated surface, due to the charge density re-distribution. If we revisit the surface with one Se vacancy, TSSs still locate inside the first QL and mixed with the dangling bond states. This can explain the dramatic modification of the Dirac cone by the vacancy in Fig. 2c.

**4 Conclusions** Selenium vacancies on the $Bi_2Se_3$ surface can modify the topological surface states, but not remove them. With a small amount of vacancies, the topological states residence inside the first QL and interact with the defect states dramatically. With a large number of vacancies or even a Bi-terminated surface, the topological states are pushed down to the second QL and consequently get separated from surface defects, preserving the original Dirac cone type of energy dispersion. This may explain the fact that ARPES usually observe a single Dirac cone similar to a perfect surface, even though lots of vacancies exist on the surface. Such kind of re-distribution of topological surface states may also be generalized to different surface defects for $Bi_2Se_3$, $Bi_2Te_3$ and $Sb_2Te_3$. In addition, we proposed to distinguish the Bi-terminated surface from Se-terminated one by STM images in experiments.

**Acknowledgements** We acknowledge the funding support by the ERC Advanced Grant (291472) and the computing time at HLRN Berlin/Hannover (Germany).

**References**

[1] X. L. Qi and S. C. Zhang, Rev. Mod. Phys. 83, 1057 (2011).
[2] M. Z. Hasan and C. L. Kane, Rev. Mod. Phys. 82, 3045 (2010).
[3] H. Zhang, C. X. Liu, X. L. Qi, X. Dai, Z. Fang, and S. C. Zhang, Nature Phys. 5, 438 (2009).
[4] Y. Xia *et al.*, Nature Phys. 5, 398 (2009).
[5] Y. L. Chen *et al.*, Science 325, 178 (2009).
[6] B. Yan and S. C. Zhang, Rep. Prog. Phys. 75, 96501 (2012).
[7] Y. S. Hor *et al.*, Phys. Rev. B 79, 195208 (2009).
[8] P. Cheng *et al.*, Phys. Rev. Lett. 105, 076801 (2010).
[9] X. He, Z. Wang, J. Shi, J. Yarmoff, APS March Meeting 2012, abstract ID: BAPS.2012.MAR.V31.4.
[10] C.X. Liu *et al.*, Phys. Rev. B 81, 041307 (2010)
[11] J. P. Perdew, K. Burke, and M. Ernzerhof, Phys. Rev. Lett. 77, 3865 (1996).
[12] G. Kresse and J. Hafner, Phys. Rev. B 47, 558 (1993).
[13] J. Tersoff and D. R. Hamann, Phys. Rev. Lett. 50, 1998 (1983).
[14] M. Koleini, T. Frauenheim, B. Yan, arXiv:1109.4000 (2012).
[15] Y.-L. Wang *et al.,* Phys. Rev. B 84, 075335 (2011).
[16] J. Honolka *et al.*, Phys. Rev. Lett. 108, 256811 (2012).
[17] W. Zhang, *et al.*, New J. Phys. 12 065013 (2010).